\documentstyle[prl,aps,preprint,epsf]{revtex} \baselineskip 0.9truecm
\large  
\begin{document}
\draft

\title{Directed lines in sparse potentials} \author{T. J. Newman and
  A. J. McKane } \address{Department of Theoretical Physics,\\ 
  University of Manchester,\\ Manchester, M13 9PL, UK} \maketitle
\begin{abstract}
  We present a continuum formulation of a $(d+1)$-dimensional directed
  line interacting with sparse potentials (i.e. $d$-dimensional
  potentials defined only at discrete longitudinal locations.) An
  iterative solution for the partition function is derived. The
  impulsive influence of the potentials induces discontinuities in the
  evolution of the probability density $P({\bf x},t)$ of the directed
  line. The effects of these discontinuities are studied in detail for
  the simple case of a single defect. We then investigate sparse
  columnar potentials defined as a periodic array of defects in
  $(2+1)$ dimensions, and solve exactly for $P$. A non-trivial
  binding/unbinding transition is found.
\end{abstract}
\vspace{5mm} 

\newpage

\section{Introduction}

The physics of directed lines (or directed polymers) has been the
focus of much interest over the past two decades. This is mainly due
to the very wide range of applicability of these simple models to
important physical processes, such as wetting\cite{wet}, the motion of
domain walls in magnets\cite{mag}, and the physics of flux lines in
superconductors\cite{sc}. The interest has been intensified by
attempting to understand the effect of disordered potentials in such
systems. In superconductors for instance, it is well known that point
disorder can help to localise the flux lines, hence allowing true
superconducting current flow; and recently, it has been realized that
columnar defects provide an even better mechanism for
pinning\cite{col}. At the model level, the addition of disorder allows
non-obvious connections to be made to other physical systems such as
spin glasses\cite{sg}, non-equilibrium interface growth\cite{kpz} and
shock waves in fluids (in terms of the noisy Burgers
equation\cite{burg}.)  A quantitative understanding of the effect of
such types of disorder on directed lines is still lacking.

Aside from these important applications, the model of a single
directed line interacting with an external potential is of substantial
interest in its own right. There has been a great deal of work on the
purely theoretical front in trying to understand these systems, with
approaches ranging from lattice RSOS descriptions\cite{wet}, to
powerful renormalization group studies\cite{rg}, and phenomenological
scaling arguments\cite{fish}.  As indicated above, what is still
lacking is a systematic way of treating strongly disordered
potentials; although much progress may be made in (1+1)
dimensions\cite{wet,oned}.  The overwhelming difficulties present in
the analytic study of directed polymers may be countered to some
degree by simplifying some aspects of the problem without trivializing
the physics. One possibility is to simplify the form of the external
potential. For instance, the problem of bulk disorder is of enormous
interest, but completely intractable above (1+1) dimensions. One may
then consider simpler scenarios (such as a very dilute limit of point
defects, or columnar disorder) where controlled analytic calculations
may provide precious insight. In fact, such is the richness of the
physics of directed lines, that even for the simple example of a
non-disordered columnar potential, there exists non-trivial behaviour
of the line, especially as a function of spatial
dimension\cite{rg,lip,new}.

In this paper we shall introduce a certain class of models which are
best described as: `a directed line interacting with sparse
potentials'. To clarify this, consider a directed line in
$(d+1)$-dimensions. We label the transverse directions by a position
vector ${\bf y}$ and the longitudinal direction by a scalar $s$. The
line generally exists in the presence of a potential $V({\bf y},s)$.
Our use of the term `sparse potentials' corresponds to the following
form for $V$:
$$V({\bf y},s) = \sum \limits _{n=1}^{\infty} v_{n}({\bf y}) \delta
(s-\tau _{n}) \ , $$ where at this level of discussion the positions
$\lbrace \tau _{n} \rbrace$ of the `impulses' $v_{n}$, along with the
functional form of the impulses themselves are left free. Specific
choices for these quantities may be made which then correspond to
physically realizable systems. For instance, taking the $\lbrace \tau
_{n} \rbrace$ to be regularly spaced leads one to consider a directed
line interacting with a set of layer potentials, as might be found in
a regular crystal. Alternatively, one may take the $\lbrace \tau _{n}
\rbrace$ to be drawn from some distribution function, which along with
taking $v_{n}({\bf y}) \sim \delta ({\bf y}-{\bf y_{n}})$, corresponds
to a directed line interacting with a very dilute set of point
defects, as may be realized in an `almost-pure' superconducting single
crystal.

Towards the end of the paper, we shall be interested in a specific
sub-class of sparse potentials, namely periodic columnar arrays of
point defects, where exact solutions are possible in the physically
relevant case of $(2+1)$-dimensions. Before such specialization, we
concern ourselves with a more general analysis of the continuum theory
of sparse potentials. An important point is that the sparse potentials
have an impulsive action on the probability density of the directed
line, leading to discontinuities in this function. Along with this
effect, is the physical constraint that the discontinuity in the
density must never be such as to make the density negative. The
short-scale consistency of the theory at these impulses will be seen
to have dramatic effects on the global properties of the probability
density of the line.

The outline of the remainder of the paper is as follows. In the next
section, we provide a general continuum formulation of the problem
following the standard methods of a path-integral description.
Section 3 is concerned with constructing an iterative solution for the
partition function for a given set of sparse potentials, due account
being taken of the discontinuous nature of this function. In section
4, we concentrate on the simple case of a single sparse potential
which is taken to be very short-ranged in the transverse directions
(i.e. approaching a delta-function.)  This corresponds to a single
point defect, and is exactly solvable. Simple as it may be, this case
reveals the subtlety of the discontinuous nature of the probability
density and allows us some intuition into the qualitative difference
of effects between attractive defects (whose effect may be global) and
repulsive defects (whose effect is always local.)  In sections 5 and 6
we concentrate on the case of a periodic array of potentials, each of
which is taken to be short-ranged -- this is essentially a columnar
array of point defects.  Section 5 sets out the solution to the case
of all the defects being either attractive or repulsive, whilst
section 6 is concerned with the solution to the case of alternating
positive and negative defects. We restrict our attention to the
physically relevant case of $(2+1)$-dimensions, which by good fortune
is the most analytically tractable. Comparison is made between our
results for these models, and recent investigations of related
microscopic models\cite{d1,d2}.  We end the paper with our conclusions
and a discussion of extensions to the present work.

\section{Formulation of the model}

As mentioned in the Introduction, there are a number of model
descriptions of directed lines, with the main difference being whether
one chooses to work in the continuum or on a lattice. In this paper we
shall use a continuum formulation which has the advantage that derived
results will not be dependent upon microscopic parameters, and thus
may be hoped to have some universal applicability.

We consider a directed line in a $(d+1)$-dimensional space, with ${\bf
  y}$ labelling the $d$ transverse directions, and $s$ labelling the
longitudinal direction. We demand that the line begins at the point
$({\bf 0},0)$ and ends at $({\bf x},t)$.  The restricted partition
function for the line, in the presence of a potential $V({\bf y},s)$
is given by
\begin{equation}
\label{part}
Z({\bf x},t) = \int \limits _{{\bf y}(0)={\bf 0}} ^{{\bf y}(t)={\bf
    x}} {\cal D}{\bf y} \ \exp \left \lbrace -\frac{1}{T} \int \limits
_{0}^{t} ds \ \left \lbrack \kappa \left ( {d{\bf y}\over ds} \right
)^{2} + V({\bf y}(s),s) \right \rbrack \right \rbrace \ ,
\end{equation}
where $T$ represents temperature, and $\kappa $ is the elastic
constant for the line.

Following standard methods\cite{schul} we may rewrite this
path-integral in the form of a partial differential equation (PDE).
Explicitly one finds
\begin{equation}
\label{pde}
\partial _{t}Z({\bf x},t) = \nu \nabla ^{2} Z - V({\bf x}, t)Z \ ,
\end{equation}
with boundary condition $Z({\bf x},0) = \delta ^{d}({\bf x})$. We have
defined a diffusion constant $\nu = T/4\kappa $, and have scaled $V$
so as to absorb a factor of $1/T$.

At this point we specialize to the system of interest, i.e. the case
where $V$ represents a set of sparse potentials. Explicitly we write
\begin{equation}
\label{pot}
V({\bf y},s) = \sum \limits _{n=1}^{\infty} v_{n}({\bf y}) \delta
(s-\tau _{n}) \ ,
\end{equation}
where $\lbrace \tau _{n} \rbrace$ represent the longitudinal locations
of the potentials $\lbrace v_{n}({\bf y}) \rbrace$.  As a final step
we integrate (\ref{pde}) using the Green function of the diffusion
equation $g({\bf x},t) = (4\pi \nu t)^{-d/2}\exp (-x^{2}/4\nu t)$,
which yields an integral equation for $Z$ of the form
\begin{equation}
\label{inteq}
Z({\bf x},t) = g({\bf x},t) - \int d^{d}x' \ \int \limits _{0}^{t} dt'
\ g({\bf x}-{\bf x}', t-t') \sum \limits _{n=1}^{\infty} v_{n}({\bf
  x}') \delta (t'-\tau _{n}) \ Z({\bf x}',t') \ .
\end{equation}

It appears to be a simple matter to integrate over the variable $t'$
using the delta-functions, but as we shall see in the next section,
this must be done with some care.

We end this section with a discussion of the physical quantities which
one can obtain from $Z$. It is important to realize that the
restricted partition function itself is not a physical quantity.  In
order to be meaningful, it must be normalized. We therefore construct
the probability density of the directed lines via
\begin{equation}
\label{prob}
P({\bf x},t) = {Z({\bf x},t) \over \int d^{d}x' Z({\bf x}',t)} \ .
\end{equation}
For future convenience we denote by $P_{0}({\bf x},t)$, the
probability density of a directed line in the absence of any external
potential (which is actually equal to the diffusion equation Green
function.)  One may also define a local `free energy' via
\begin{equation}
\label{free}
f({\bf x},t) = -T \log (Z({\bf x},t)) \ .
\end{equation}
We use the term `free energy' guardedly, since this quantity, as
defined above, is not extensive (in terms of the length $t$ of the
lines). It is, however, a useful measure of the energy/entropy balance
for a given end-point value ${\bf x}$. For instance, if one chooses
$V$ to be a columnar potential (which exists only for ${\bf x}={\bf
  0}$), then $f({\bf 0},t)$ is a sensitive measure of a bound line ($f
\sim t$), as opposed to an unbound line (generally $f \sim \log(t)$.)
In this paper we shall generally intuit the behaviour of the line from
studying the probability density.

As a final remark, we stress that although the path-integral and
partial differential equation forms for $Z$ bear a striking
resemblance to the Feynman path-integral and Schr\"odinger equation
descriptions of a quantum mechanical particle respectively; one should
use caution in applying results from quantum mechanics to the present
problem -- the idea that the problems are related by a simple Wick
rotation ($\tau = it$) is over-simplified. Physically, the partition
function as defined above is a positive-definite quantity.  There is
no analogous constraint on the complex wave function of quantum
mechanics. Also, the time evolution of the wave function is generally
postulated to be continuous\cite{ll}.  In the above problem, there is
no physical constraint on the continuity of the partition function as
a function of $t$. The reason for this difference is the following. In
quantum mechanics a discontinuity in the time evolution of the wave
function corresponds to a temporal discontinuity in the probability
density of the quantum mechanical particle, which is physically
unreasonable. In contrast, the discontinuity of the partition function
breaks no physical laws. This is because the symbol $t$ used above
corresponds to the {\it length of the line} - this is a fixed quantity
for a given line. The meaning of the rate of change of the partition
function with respect to $t$ is to take two lines whose lengths are
$t$ and $t+\delta t$ respectively, and to equilibrate them in
identical thermal (and possibly quenched disorder) environments;
thereafter measuring the partition function for each line. There is no
reason {\it a priori} to insist on continuity of the partition
function.

These general distinctions between quantum mechanics (as expressed by
Feynman's path integral) and the statistical mechanics of a directed
line (as expressed by (\ref{part}) above) will have important
consequences in the sequel of this paper.

\section{Discontinuities and `General Solution'}

To gain some insight into the nature of a sparse potential, let us
simplify the problem to having a single potential $v({\bf x})$ located
at a longitudinal position $\tau $.  The integral equation for $Z$ now
takes the form
\begin{equation}
\label{inteq1}
Z({\bf x},t) = g({\bf x},t) - \int d^{d}x' \ \int \limits _{0}^{t} dt'
\ g({\bf x}-{\bf x}', t-t') v({\bf x}') \delta (t'-\tau ) \ Z({\bf
  x}',t') \ .
\end{equation}
Obviously, for $t<\tau$ we have the solution $Z({\bf x},t) = g({\bf
  x},t)$ which implies $P({\bf x},t) = P_{0}({\bf x},t) = g({\bf
  x},t)$.  In particular, $Z^{-}({\bf x}) \equiv \lim \limits
_{\epsilon \rightarrow 0} Z({\bf x}, \tau -\epsilon) = g({\bf x},\tau
)$. We now go on to the more difficult task of finding $Z^{+}({\bf x})
\equiv \lim \limits _{\epsilon \rightarrow 0} Z({\bf x}, \tau
+\epsilon)$.

In taking the limit $t \rightarrow \tau$ care is needed when
integrating over the delta-function in (\ref{inteq1}), but the Green
function $g({\bf x}-{\bf x}', t-t') = g({\bf x}-{\bf x}', t-\tau )$
may safely be replaced by $\delta ({\bf x}-{\bf x}')$ leading to
\begin{equation}
\label{inteq11}
Z^{+}({\bf x}) = g({\bf x},\tau ) - v({\bf x}) \lim \limits _{\epsilon
  \rightarrow 0} \int \limits _{0}^{\tau + \epsilon} dt' \delta
(t'-\tau ) \ Z({\bf x},t') \ .
\end{equation}
Therefore $Z^{+}({\bf x}) = g({\bf x},\tau ) + O(v)$, and since this
also holds for $Z^{-}({\bf x})$ to this order, we may write
(\ref{inteq11}) as
\begin{eqnarray}
\label{inteq12}
Z^{+}({\bf x}) & = & g({\bf x},\tau ) - v({\bf x}) \lim \limits
_{\epsilon \rightarrow 0} \int \limits _{0}^{\tau + \epsilon} dt'
\delta (t'-\tau ) \ g({\bf x}, \tau ) \ + O(v^2) \ \nonumber \\ & = &
g({\bf x},\tau )\left[ 1 - v({\bf x}) + O(v^2) \right] \ .
\end{eqnarray}
It is important to note that, while one can expand the function $g$
about $t = \tau$, this is not so for $Z$, which can now clearly be
seen to be discontinuous at this point.

To find the higher order terms in (\ref{inteq12}), we write
$Z^{+}({\bf x})$ and $Z^{-}({\bf x})$ in the unified form:
\begin{equation}
\label{inteq13}
Z({\bf x},t) = g({\bf x},\tau ) - v({\bf x}) \ \int \limits _{0}^{t}
dt' \delta (t'-\tau ) \ Z({\bf x},t') \ .
\end{equation}
where $t = \tau \pm \epsilon , \epsilon \rightarrow 0$. Iterating this
equation gives
\begin{equation}
\label{inteq14}
Z({\bf x},t) = g({\bf x},\tau )\left\{1 - \ \int \limits _{0}^{t} dt'
v \delta (t'-\tau ) + \ \int \limits _{0}^{t} dt' \int \limits
_{0}^{t'} dt'' v^2 \delta (t'-\tau ) \, \delta (t''-\tau ) \ + \ldots
\right\} \ .
\end{equation}
For $t<\tau$ the integrals give no contribution, confirming the result
for $Z^{-}({\bf x})$, and for $t>\tau$ the lower limits of the
integrals may be taken to minus infinity and $t$ replaced by plus
infinity, giving
\begin{equation}
\label{inteq15}
Z^{+}({\bf x}) = g({\bf x},\tau )\left\{1 - \ \int \limits
_{-\infty}^{\infty} dt' v \delta (t'-\tau ) \ + \frac{1}{2} \int
\limits _{-\infty}^{\infty} dt' \int \limits _{-\infty}^{\infty} dt''
v^2 \delta (t'-\tau ) \, \delta (t''-\tau ) \ + \ldots \right\} \ ,
\end{equation}
where the symmetry of the double integral has been used to rewrite it
as an integral over all $t$ and $t'$, hence the factor of $1/2$.
Continuing this procedure one finds that
\begin{equation}
\label{inteq16}
Z^{+}({\bf x}) = g({\bf x},\tau ) \ \sum_{n=0}^{\infty} \frac{1}{n!} \ 
\left\{ - \int^{\infty}_{-\infty} dt' \, v \delta (t' -\tau )
\right\}^{n} \ .
\end{equation}
Therefore, in the limit $\epsilon \rightarrow 0$,
\begin{equation}
\label{inteq17}
Z({\bf x},t) = \left\{ \begin{array}{ll} g({\bf x},\tau )\, \exp (-
v({\bf x})) , & \mbox{\ if $t = \tau + \epsilon$} \\ g({\bf x},\tau ),
& \mbox{\ if $t = \tau - \epsilon$}
\end{array} \right.
\end{equation}

From (\ref{inteq13}) it is clear that, in the limit $t \rightarrow
\tau$, the coordinate $x$ is simply a label and plays no significant
part in the phenomenon we have just highlighted. Therefore, some
further insight into this effect may be gained by a study of the
zero-dimensional versions of these models. In order to do this it is
useful to examine the differential equations corresponding to the
above integral equations. The one corresponding to (\ref{inteq1}) has
the form (cf. (\ref{pde})):
\begin{equation}
\label{pde1}
\partial _{t}Z({\bf x},t) = \nu \nabla ^{2} Z - v({\bf x}) 
\delta (t-\tau) \ Z \ .
\end{equation}
The zero-dimensional version of this equation is
\begin{equation}
\label{ode1}
\partial _{t}Z(t) = - v \, \delta (t-\tau) \ Z \ .
\end{equation}
It is easy to check that the integral equation corresponding to
(\ref{ode1}) is simply (\ref{inteq13}) but with the $x$ label absent.
On the other hand, we can in this simple case, solve (\ref{ode1})
directly to find:
\begin{equation}
\label{ode2}
Z(t) = \left\{ \begin{array}{ll} g(\tau )\, \exp (-v) , & \mbox{\ if
  $t = \tau + \epsilon$} \\ g(\tau )\, \exp (-v\theta (0)) , & \mbox{\ 
  if $t = \tau$} \\ g(\tau ), & \mbox{\ if $t = \tau - \epsilon$}
\end{array} \right.
\end{equation}
in the limit $\epsilon \rightarrow 0$. We conclude that, while one may
meaningfully formulate questions about the discontinuity of $Z$ at
$t=\tau$, the definition of the partition function is itself
ill-defined precisely at this point, depending as it does on the
definition of $\theta (0) \equiv \int^{0}_{-\infty} dt \delta (t)$. Of
course, in a microscopic approach, the delta-function would be
smoothed out and this ambiguity would be absent. However, as far as
the evolution of $Z$ is concerned, only its change at $t=\tau$, and
not its actual value there, is relevant. So the conclusions that we
draw in this paper will be independent of the precise form of any
underlying microscopic model.

Returning to the general form of the potential (\ref{pot}), the
relation
\begin{equation}
\label{discon}
Z^{+}({\bf x},\tau ) = Z^{-}({\bf x},\tau )\, \exp (-v({\bf x})) \ ,
\end{equation}
allows us to write a general iterative solution for the partition
function.  The idea is to split the evolution of $Z$ into two parts;
the first being concerned with the change in $Z$ as the line
encounters a sparse potential, the second with the evolution of $Z$
between potentials. We naturally define
\begin{equation}
\label{def1}
Z^{-}_{n}({\bf x}) = \lim \limits _{\epsilon \rightarrow 0} Z({\bf x},
\tau _{n}-\epsilon) \ ,
\end{equation}
and also
\begin{equation}
\label{def2}
Z^{+}_{n}({\bf x}) = \lim \limits _{\epsilon \rightarrow 0} Z({\bf x},
\tau _{n}+\epsilon) \ .
\end{equation}
Directly using eqn. (\ref{discon}) we have
\begin{equation}
\label{dc2}
Z^{+}_{n}({\bf x}) = Z^{-}_{n}({\bf x})\, \exp (-v_{n}({\bf x})) \ ,
\end{equation}
The evolution of the partition function between potentials is easily
obtained since it is nothing but thermal wandering. We therefore have
\begin{equation}
\label{freew}
Z^{-}_{n+1}({\bf x}) = \int d^{d}x' \ g({\bf x}-{\bf x}', \tau _{n+1}
- \tau _{n}) \ Z^{+}_{n}({\bf x}') \ .
\end{equation} 

Eqns. (\ref{dc2}) and (\ref{freew}) are the main results of this
section and constitute an iterative solution for the partition
function, in some ways analogous to the usual transfer matrix solution
used in discrete lattice formulations of directed walks\cite{wet}.
Once the set of functions $Z^{+}_{n}$ and $Z^{-}_{n}$ is known, the
partition function at intermediate values of the longitudinal
co-ordinate may be found by quadrature from (\ref{inteq}).

\section{Single Defect}

In this section we solve perhaps the simplest example of a sparse
potential, namely a short-ranged potential corresponding to a single
point defect, located at longitudinal location $s=\tau$.  For
convenience we choose $v({\bf y}) = -\rho \Delta ({\bf y})$, where
\begin{equation}
\label{srp}
\Delta ({\bf y}) = (\pi a^{2})^{-d/2} \ \exp (-y^{2}/a^{2}) \ .
\end{equation}
The scale $a$ is to be regarded as the shortest transverse scale in
the problem, although we shall always need to keep $a$ non-zero in
order to regularize the theory. [Note, in the limit of $a \rightarrow
0$, the function $\Delta $ becomes a Dirac delta-function.] The
parameter $\rho $ simply represents the strength of the `defect' --
for $\rho > 0$ the defect is attractive, whilst for $\rho < 0$, the
defect is repulsive.  We can now go on to calculate the probability
densities.

Since the only potential in the system is that due to the single
defect, we clearly have $Z^{-}({\bf x},\tau ) = g({\bf x},\tau )$
which implies $P^{-}({\bf x},\tau ) = P_{0}({\bf x},\tau )$.
Integrating $Z^{+}({\bf x},\tau )$, as given in (\ref{inteq17}), over
${\bf x}$ in order to find the appropriate normalization leads us to
an expression for the probability density of the line on the positive
side of the defect.  It is convenient when discussing this quantity to
introduce a length $l = (4 \nu \tau)^{1/2} \gg a$, which is the
effective transverse wandering of the line between the origin and the
defect, and then to form appropriately scaled versions of $\rho$ and
$l$ by defining $\rho ^{*} = \rho /(\pi a^2 )^{d/2}$ and $l^{*} =
l/a$. One then finds that
\begin{equation}
\label{relprob}
P^{+}({\bf x},\tau) = P_{0}({\bf x}, \tau) \ \left \lbrace {
  (l^{*})^{d} \exp \rho \Delta ({x}) \over e^{\rho ^{*}} + (l^{*})^{d}
  - 1 } \right \rbrace
\end{equation}
Looking at the short-range form of this expression it is immediately
clear that
\begin{equation}
\label{srprob}
{P^{+}({\bf 0},\tau ) \over P_{0}({\bf 0},\tau )} \sim \left \lbrace {
  (l^{*})^{d}\, e^{\rho ^{*}} \over e^{\rho ^{*}} + (l^{*})^{d} - 1 }
\right \rbrace
\end{equation}
whilst for $|{\bf x}| \rightarrow \infty $ we have
\begin{equation}
\label{lrprob}
{P^{+}({\bf x},\tau ) \over P_{0}({\bf x},\tau )} \sim \left \lbrace {
  (l^{*})^{d} \over e^{\rho ^{*}} + (l^{*})^{d} - 1 } \right \rbrace
\end{equation}

Examination of these expressions for the relative discontinuity of the
probability density reveals the following effects:

\noindent{\bf Attractive weak defect}\ $\rho > 0$ and $\rho = O(a^d)$

\medskip

In this case $\rho ^{*} = O(1)$, and since $l^{*} \gg 1$, the
short-range discontinuity of the probability density (as given by
(\ref{srprob})) is of order unity, whilst the long-range discontinuity
(as given by (\ref{lrprob})) is of negligible size.

\medskip

\noindent{\bf Attractive strong defect}\ $\rho > 0$ and $\rho = O(1)$

\medskip

The situation is markedly different here: for small $|{\bf x}|$,
$P^{+} \gg P_{0}$ and for large $|{\bf x}|$, $P^{+} \ll P_{0}$. The
two probability densities become equal at some critical value of
$|{\bf x}|$ which is much smaller than $a$.

\noindent{\bf Repulsive defect}\ $\rho < 0$

\medskip

The conclusions for weak repulsive defects are exactly as for weak
attractive defects. For strong repulsive defects $P^{+} \ll P_{0}$ at
short range and $P^{+} \approx P_{0}$ at long range.

\medskip

These results can be summarized by saying that at long range (in
practice for $|{\bf x}|>a$) weak attractive defects and all repulsive
defects have no effect. But a strong attractive defect does have a
global effect on the probability density of the directed line, at the
longitudinal location $\tau$.

We shall briefly consider the form of the probability density for
$t>\tau$.  To do this we write down an integral equation of the form
(\ref{inteq1}), but with initial time $\tau _{+}$. Using $Z^{+}({\bf
  x},\tau )$, as given by (\ref{inteq17}), we find:
\begin{equation}
\label{evop}
{P({\bf x},t) \over P_{0}({\bf x},t) } \approx {(l^{*})^{d} + F({\bf
    x},t)\, ( e^{\rho ^{*}} - 1 ) \over (l^{*})^{d} + e^{\rho ^{*}} -
  1}
\end{equation}
where
\begin{equation}
\label{eff}
F({\bf x},t) = \frac{1}{\left( 1 - \gamma \right)^{d/2}}\, \exp \left
[- \ \frac{{\bf x}^2}{4\nu t^2} \ \frac{\tau}{(1 - \gamma )}\right ]
\end{equation}
and where we have defined $\gamma (t) = \tau/ t $.

Let us examine the consequences of this result for times significantly
greater than $\tau $; i.e. we take $\gamma \ll 1$.  Firstly, we note
that the effect of the defect upon the probability density is
negligible, if the defect is repulsive or attractive and weak, since
the quantity $(l^{*})^d$ dominates in both the numerator and
denominator of (\ref{evop}). The situation is again more interesting
when we consider a strong attractive defect. It is now the quantity
$e^{\rho ^{*}}$ which dominates, at least for small enough $x$.
Therefore, the healing of the distribution function at ${\bf x}={\bf
  0}$ follows $(P/P_{0})_{x=0} \sim F({\bf 0},t) \sim 1 + d\tau /2t$.
For transverse distance $|{\bf x}|$ being large (actually $|{\bf x}|
\gg t(\nu/\tau)^{1/2}$), the healing does not occur except at
extremely large longitudinal distances; the relative difference in the
probability density (compared to the free case) satisfying
$(P/P_{0})_{x=\infty } \sim (l^{*})^{d}\, e^{-\rho ^{*}}$. There will
exist a scale $L(t)$ at which the ratio of $P$ to $P_{0}$ is exactly
unity, given by $F({\bf x},t) \sim 1$. From Eq.(\ref{eff}) we find
$L(t) = (2d\nu t)^{1/2}$.
 
\medskip

So, to summarize the results for the simple situation of a single
localized defect of strength $\rho$, we find that there is a
qualitative difference between attractive and repulsive defects. In
the former case, there exist two classes of defect -- weak and strong
-- which are distinguished by their effect upon the probability
density, this effect being local and global respectively. In the
latter case (repulsive defect), we find that for any strength of
defect, the effect upon the probability density is always local. The
extreme asymmetry in effect of positive and negative localized defects
will be seen to have interesting consequences in the next two sections
in which we consider infinite arrays of defects.  Since the effect of
a repulsive defect upon the line is qualitatively the same for any
strength of defect, we shall not distinguish between weak and strong
repulsive defects. In the following sections we shall generally take
the strength of the repulsive defect to be of order unity.

\section{Periodic Columnar Potential I}

In this section we shall consider a more complicated situation, namely
an infinite periodic array of localized defects located on the column
defined by ${\bf x}=0$. We henceforth restrict our attention to
(2+1)-dimensions.  This shares the attractive features of being both
the most physically interesting case, as well the most analytically
tractable -- a rare coincidence. Choosing $\tau $ to be the
longitudinal separation between the defects, and $\rho $ to be their
strength, we consider a potential of the form
\begin{equation}
\label{ppot1}
V({\bf y},s) = -\rho \sum \limits _{n=1}^{\infty} \Delta ({\bf y})
\delta (s-\tau n) \ ,
\end{equation}
where we adopt the gaussian envelope form (\ref{srp}) for the
short-range function $\Delta $. The range of the potentials is of
$O(a)$ which we take to be the smallest transverse scale in the
problem. In particular we have $a \ll \l$ where $l=(4\nu \tau)^{1/2}$
is the effective transverse wandering of the line between defects. All
results will be derived to leading order in $a$; the fact that the
transverse scale of the function $\Delta ({\bf x})$ is of $O(a)$
allows us to frequently implement it as a Dirac delta function to get
results to this order. We have chosen all the defects to have the same
strength $\rho $ which we can take to be either positive (attractive
defects) or negative (repulsive defects). In the next section we shall
study an analogous situation, but with alternating attractive and
repulsive defects.

Taking the general iterative solution as given in eqns. (\ref{dc2})
and (\ref{freew}) and substituting the explicit form for the potential
above, yields the relations
\begin{equation}
\label{sol1}
Z^{-}_{n+1}({\bf x}) = \int d^{d}x' \ g({\bf x}-{\bf x}', \tau)
Z^{+}_{n}({\bf x}') \ ,
\end{equation}
and
\begin{equation}
\label{sol2}
Z^{+}_{n}({\bf x}) = Z^{-}_{n}({\bf x}) \, \exp (\rho\Delta ({\bf x}))
\ .
\end{equation}
Combining these two results and implementing $\Delta $ as a Dirac
delta function wherever possible we find
\begin{eqnarray}
\label{first}
Z^{-}_{n}({\bf x}) & = & \frac{Z^{-}_{n-1}({\bf 0})}{\Delta ({\bf
    0})}\, g({\bf x},\tau ) \, \left[ e^{\rho\Delta ({\bf 0})} - 1
\right] + \int d^{d}x' \ g({\bf x}-{\bf x}', \tau)\, Z^{-}_{n-1}({\bf
  x'}) \nonumber \\ \nonumber \\ & = & \frac{Z^{-}_{n-1}({\bf
    0})}{\Delta ({\bf 0})}\, g({\bf x},\tau ) \, \left[ e^{\rho\Delta
  ({\bf 0})} - 1 \right] + \int d^{d}x' \ g({\bf x}-{\bf x}', 2\tau)\,
Z^{-}_{n-2}({\bf x'})\, \exp \rho\Delta ({\bf x'})
\end{eqnarray}
Repeating this procedure leads us to
\begin{equation}
\label{solution}
Z^{-}_{n}({\bf x}) = g({\bf x}, n\tau ) + R \sum \limits _{m=1}^{n-1}
g({\bf x}, (n-m)\tau ) Z^{-}_{m}({\bf 0}) .
\end{equation}
where $R$ is defined by
\begin{equation}
\label{r}
R = \frac{e^{\rho\Delta ({\bf 0})} - 1}{\Delta ({\bf 0})} \ .
\end{equation}
It is convenient at this point to define $\psi _{n} = Z^{-}_{n}({\bf
  0})$, along with $f_{n} = g({\bf 0}, \tau n) = 1/(\pi n l^{2})$.
Setting ${\bf x} = {\bf 0}$ in the above equation then gives
\begin{equation}
\label{final1}
\psi _{n} = f_{n} + R \sum \limits _{m=1}^{n-1} f_{n-m} \psi _{m} \ .
\end{equation}
This discrete equation may be solved exactly by making use of a
generating function. The details of the calculation are relegated to
Appendix A. The resulting form for $\psi _{n}$ depends on whether
defects are attractive or repulsive. Thus we consider these two cases
separately.

\subsection{Attractive defects}

From Appendix A, the asymptotic (i.e. $n \gg 1$) result for $\psi
_{n}$ takes the form
\begin{equation}
\label{asym1}
R \psi _{n} \sim \left \lbrack e^{-1/R '} \over R ' ( 1 - e^{-1/R '})
\right \rbrack \ \exp \left \lbrack n \ln \left ( {1 \over 1 - e^{-1/R
    '}} \right ) \right \rbrack \ ,
\end{equation}
where $R ' = R / \pi l^{2} = [ e^{\rho ^{*}} - 1]/(l^{*})^{2}$. So,
for any $\rho > 0$, the partition function, evaluated at a defect
site, grows exponentially; but with a rate which vanishes
exponentially fast for small $\rho $. In order to obtain the
physically meaningful probability density, we must normalize the
partition function. Generally, the normalization is defined as $N(t) =
\int d^{d}x Z({\bf x},t)$. In particular, we define
\begin{equation}
\label{defno}
N_{n} = \lim _{\epsilon \rightarrow 0} N(\tau n - \epsilon) \ .
\end{equation}
Then by integrating Eq. (\ref{solution}) over the transverse space we
obtain
\begin{equation}
\label{norm1}
N_{n} = 1 + R \sum \limits _{m=1}^{n-1} \psi _{m} \ ,
\end{equation}
which on evaluating the sum gives, for large $n$,
\begin{equation}
\label{normasy}
N_{n} \sim (1/R ') \ \exp \left \lbrack n \ln \left ( {1 \over 1 -
  e^{-1/R '}} \right ) \right \rbrack \ .
\end{equation}
Dividing the partition function (\ref{asym1}) by this normalization
(\ref{normasy}) yields the asymptotic form for the probability density
at the $n^{\rm th}$ defect:
\begin{equation}
\label{pdn}
P_{n}({\bf 0}) \equiv \lim _{\epsilon \rightarrow 0} P({\bf 0},\tau n
- \epsilon) \sim { e^{-1/R '} \over R (1 - e^{-1/R '})} \ .
\end{equation}
From this result we see that for any $\rho > 0$, the probability
density on the column (actually on a defect site) attains a non-zero,
constant value as $n \rightarrow \infty$. This indicates that the line
is always {\it bound} to the array of defects, regardless of how
weakly attractive they are, or how widely separated.

It is interesting to calculate the probability density on the column,
but in between the defects. This may be directly evaluated by making
use of eqn. (\ref{inteq}). Setting $t=\tau (n+\theta )$ with $\theta
\in (0,1]$ we find the asymptotic result
\begin{equation}
\label{probd}
P({\bf 0},t) \sim { R ' \ e^{-1/R '} \over R (1 - e^{-1/R '}) ^{\theta
    } } \ F_{\theta } \left ( - \ln(1-e^{-1/R '}) \right ) \ ,
\end{equation}
where
\begin{equation}
\label{whatisf}
F_{\theta }(p) \equiv \int \limits _{p}^{\infty} du \ {e^{-u\theta }
  \over ( 1 - e^{-u} ) } \ .
\end{equation}
On setting $\theta = 1$, the above expression reduces to the
asymptotic result for defect sites, as given by (\ref{pdn}).

In the limit of very strong attractive defects, the probability
density along the column has the form $P({\bf 0},(n+\theta)\tau) =
(\theta \pi l^{2})^{-1}$ for $0<\theta\le 1$.  The density takes its
largest value on the positive side of the defect, and then decays as
$1/\theta $ until the next defect is reached ($\theta =1$.)

We also note that in the limit of vanishing defect strength $P({\bf
  0},t)$ reduces to the same form for both defect sites, and positions
in between. Explicitly one has
\begin{equation}
\label{small}
P({\bf 0}, t) \sim e^{-(l*)^2/\rho ^{*}} / \rho \ , \ \ \ \ \ \ \ \ 
\rho ^{*} \ll 1 \ , \ t \gg \tau \ .
\end{equation}

In the sense that the array binds the line for any $\rho > 0$, we may
say that it acts in precisely the same way as a constant energy column
(CEC) which is attractive\cite{rg,lip,new}. In that case, one has a
potential of the form $V({\bf y},s) = -{\bar \rho} \Delta ({\bf y})$.
The asymptotic form of the probability density on the column, for
vanishingly small ${\bar \rho }$, takes the form
\begin{equation}
\label{smallw}
P({\bf 0}, t) \sim e^{-1/{\tilde \rho }} / a^{2}{\tilde \rho} \ ,
\end{equation}
where ${\tilde \rho} = {\bar \rho }/(4 \pi \nu)$. Comparing these
results, we see that concerning the dominant exponential behaviour,
there is an effective renormalization of the defect strength, such
that it appears as the strength of a CEC. The precise form of this
renormalization is ${\bar \rho } = \rho /\tau$ which is an intuitively
appealing result.

\subsection{Repulsive defects}

The solution to this discrete equation for the partition function is
outlined in Appendix A, along with details of the evaluation of the
normalization.  On dividing $\psi _{n}$ by $N_{n}$, we obtain the
probability density at a defect site. We find it to have the
asymptotic form
\begin{equation}
\label{probrep}
\nonumber P_{n}({\bf 0}) = \left \lbrace n \pi l^{2} (R ')^{2} \left
\lbrack \ln (n e^{1/|R '|}) \right \rbrack ^{2} \right \rbrace ^{-1} =
\left \{
\begin{array}{ll}
  \Bigl \lbrack n \pi l^{2} \Bigr \rbrack ^{-1} \ , & 1 \ll n \ll
  e^{1/|R '|} \\ \Bigl \lbrack \pi l^{2} (R ')^{2} n(\ln (n))^{2}
  \Bigr \rbrack ^{-1} \ , & \ \ \ \ n \gg e^{1/|R '|}
\end{array}
\right.
\end{equation}
So in the deep asymptotic regime, the probability density at a defect
location decays as $P_{n}({\bf 0}) \sim [n (\ln (n))^{2}]^{-1}$.
Again, it is interesting to compare this result to that obtained for
the case of a CEC (this time with repulsive energy.)  Following the
methods of ref.\cite{new} one may ascertain that for the CEC, $P({\bf
  0}, t) \sim [ t \ln(t) ]^{-1}$ in the asymptotic regime. It
therefore appears as if the defects repel the line more effectively
than a CEC; which is counter-intuitive. The situation may be clarified
by calculating the probability density in between the defects, i.e.
taking $t=\tau (n+\theta)$ with $n \gg 1$.  This may be done by making
use of (\ref{inteq}) with the result that
\begin{equation}
\label{ppp}
P({\bf 0},t) \sim {1 \over t} \left \lbrack 1 + O \left ( { 1\over \ln
  (t) } \right ) + O \left ( { 1 \over \theta (\ln(t))^{2} } \right )
\right \rbrack \ .
\end{equation}
This is the asymptotic behaviour of a free line. Therefore the
repulsive array has no qualitative effect upon the probability density
except right at the defect positions.  The decay of $P({\bf 0},t)$ for
the CEC is marginally faster than a free line, but marginally slower
than that of a line constrained to pass through a defect -- this is an
intuitively acceptable result.  In contradistinction to the case of
attractive defects, there is no effective renormalization of the
defects into a CEC when they are repulsive.

\section{Periodic Columnar Potential II}

In the last section we have seen that there exists a great difference
between an infinite array of attractive defects, and an infinite array
of repulsive defects. In the former case, the line is bound to the
array; and for small values of the potential energy, the array acts
precisely in the same manner as a CEC. In the latter case, the line is
oblivious to the column on which the array is defined, except directly
at defect sites. In that case, the probability density is marginally
reduced. There is no relation between the repulsive array, and a
repulsive CEC. One may ask how the line acts when the array consists
of both attractive and repulsive defects. In fact our initial
motivation was to study the case of a random admixture of such defects
along the column. However, this simple example of a disordered
potential is extremely difficult to analyse. A simpler task is to
arrange the attractive and repulsive defects in an alternating pattern
along the column. The physics of such a system has attracted much
interest recently \cite{d1,d2}, albeit in a microscopic formulation in
terms of the RSOS model.  Our presentation in this section will be
rather brief as most of the calculation may be constructed using the
methods of the past two sections.  Also, we shall content ourselves
with only examining the gross features of this system; namely the
location of the binding/unbinding transition, and the qualitative
behaviour of the critical properties of the line and the bound phase.

Denoting the strength of the attractive defects by $\rho > 0$, and
that of the repulsive defects by $-\sigma < 0$, we consider the set of
sparse potentials
\begin{equation}
\label{ppot2}
V({\bf y},s) = -\rho \sum \limits _{n=1}^{\infty} \Delta ({\bf y})
\delta (s-2n\tau) + \sigma \sum \limits _{n=1}^{\infty} \Delta ({\bf
  y}) \delta (s-(2n-1)\tau ) \ .
\end{equation}
Following a similar procedure to that used in the previous section we
may derive a closed equation for the partition function at a defect
site. In this case there is the minor complication of having two types
of defect, which may be easily accommodated in the following way. We
denote the partition function at attractive and repulsive defect sites
by $\psi _{n}^{E}$ and $\psi _{n}^{O}$ respectively: $\psi _{n}^{E}
\equiv Z^{-}_{2n}({\bf 0})$ and $\psi _{n}^{O} \equiv
Z^{-}_{2n-1}({\bf 0})$. To derive the iterative equation for these
quantities, one uses the fundamental relations (\ref{dc2}) and
(\ref{freew}) and follows a similar procedure to that described in the
previous section. This leads to the following simultaneous equations:
\begin{eqnarray}
\label{iteo}
\nonumber \psi _{n}^{E} & = & f_{2n} + R \sum \limits
_{m=1}^{n-1}f_{2n-2m} \psi _{m}^{E} - S \sum \limits _{m=1}^{n}
f_{2n-2m+1} \psi _{m}^{O} \\ \psi _{n}^{O} & = & f_{2n-1} + R \sum
\limits _{m=1}^{n-1} f_{2n-2m-1} \psi _{m}^{E} - S \sum \limits
_{m=1}^{n-1} f_{2n-2m} \psi _{m}^{O} \ ,
\end{eqnarray}
where $R$ is given by (\ref{r}) and
\begin{equation}
\label{s}
S = \frac{1 - e^{-\sigma\Delta ({\bf 0})}}{\Delta ({\bf 0})} \ .
\end{equation}

We shall relegate the explicit solution of these equations to Appendix
B.  The main result to emerge from this solution is the shifting of
the binding/unbinding transition to a critical line in the $(\rho ,
\sigma )$ plane which is illustrated in fig.1. The precise equation
for this line is given in (\ref{crit}), but the general structure
takes the form
\begin{equation}
\label{critcond}
\sigma = \left\{ \begin{array}{ll} \rho \ , & \mbox{\ if $\rho \ll
  \rho _{c} $} \\ -\pi a^{2} \ln \left( \frac{\rho _{c} - \rho}{\pi
  a^2} \right)\ , & \mbox{\ if $\rho \rightarrow \rho _{c} $}
\end{array} \right.
\end{equation}
where $\rho _{c} = \pi a^{2}\ln 2$. It is interesting to note that for
$\rho > \rho_{c}$ the line is always bound, regardless of the strength
of the repulsive sites.  This result is in qualitative agreement with
the recent work from the Fribourg group \cite{d1,d2}.  This physics
may have been intuitively expected following the analysis of the
single defect, where the vast difference of effect between attractive
and repulsive defects was examined in detail.  In the terminology of
section 4, we may say that strong attractive defects will always bind
the line, whereas weak attractive defects require a critical strength
in order to do this. There is no analogy of `weak' and `strong' for
repulsive defects.

Some further details are examined in Appendix B which we shall
summarize here. Firstly one may examine the behaviour of the line at
criticality. In this case one finds that the probability density of
the line at a defect site (attractive or repulsive) follows the
asymptotic behaviour of a free line, namely $P({\bf 0}, n\tau) \sim
1/n$. One may also study the bound state in which case one finds that
the probability density saturates to a constant at both attractive and
repulsive defect sites, although the density is a factor of
$(1/l^{*})^{2}$ smaller at the repulsive defect sites. One then has the
picture that the although the line is bound, it really binds only to
the attractive defects, and has a vanishingly small probability
density at the repulsive defect sites.  Having gleaned the main
qualitative aspects of the alternating column, we shall end this
section here and proceed to presenting our conclusions.

\section{Conclusions}
In this paper we have examined a class of models described as `a
directed line in the presence of sparse potentials'; with the
understanding that a sparse potential is a $d$-dimensional potential
defined at a single longitudinal location. In section 3 we obtained a
general iterative solution for the partition function which consisted
of two pieces: free propagation between potentials, and a
discontinuity when passing through a potential. In section 4 we
considered in some detail a single short-ranged sparse potential which
corresponds to a single defect. The probability density of the line
was evaluated for both attractive and repulsive defects. In the former
case, the density was seen to undergo a global discontinuous change
for strong defects, and the healing of the density (i.e. the
relaxation to the density of a free line) was found to be incomplete
for arbitrarily large longitudinal distances above the defect. The
latter case of a repulsive defect was completely different in that for
any strength of defect, the density of the line is undisturbed except
within a small region about the defect.  [As a brief aside we may
relate this extreme asymmetry to the behaviour of a non-equilibrium
interface evolving under the Kardar-Parisi-Zhang (KPZ)
equation\cite{kpz}.  Under the mapping between directed lines and the
KPZ equation, an attractive (repulsive) defect corresponds to an
upward (downward) force, acting for a short duration upon the surface.
It has been previously observed\cite{new,newb} that in the
strong-coupling regime of the KPZ equation, an upwards force of
sufficient strength may seed a large disturbance in the interface
which then becomes effectively frozen. Alternatively, a downward force
of arbitrarily large strength plays no role, since any disturbance it
causes is quickly eradicated by the strong upward action of the KPZ
non-linearity. The behaviour of the directed line under the influence
of a single defect is seen to exhibit an analogous effect.  The
possibility of gaining intuition concerning the strong-coupling
behaviour of the KPZ equation is a prime example of the usefulness of
studying non-trivial, yet tractable, directed line models of the type
considered in this paper.]

In section 5 we studied a periodic array of defects arranged on a
column, exclusively in (2+1)-dimensions, which is of most interest.
For attractive defects, the line was found to be asymptotically bound,
and indeed, the saturated from of the probability density along the
column (for vanishingly weak potentials) was found to correspond to
that obtained previously for a constant energy column (CEC),
indicating that the line samples the defects in such a way as to
renormalize their effect to be that of a CEC. For an array consisting
of repulsive defects, we found that the density along the column is
qualitatively unchanged from that of a free line, i.e.  $P \sim 1/t$.
This result is logarithmically modified at the defect positions,
having the form $P \sim 1/(t \ln ^{2}(t))$. There is no relation of
these results to a repulsive CEC where one has $P \sim 1/(t \ln (t))$,
indicating that the sampling of the defects by the line does not have
a renormalizing, or smoothing effect. [The fact that the line may be
bound by an array of attractive defects has a novel implication for
the KPZ equation; namely, that a sequence of discrete upward impulses
is sufficient to move the interface with non-zero velocity, similar to
the effect of pushing with a constant force\cite{new}.] In the last
section we examined an array consisting of alternating attractive
(with strength $\rho > 0$) and repulsive (strength $-\sigma < 0$)
defects. It was found that the binding/unbinding transition is shifted
to a location in the $(\sigma, \rho)$ phase plane defined by the
condition $\sigma = \rho /(1-\rho/\rho _{c})$. This result is
interesting as it indicates that for attractive defects of strength
greater than the critical strength $\rho _{c} = \pi a^{2}\ln (2)$, the
line will always be bound, regardless of the strength of the repulsive
defects.  This latter result is in accord with recent calculations on
an equivalent microscopic RSOS model\cite{d1,d2}.  The asymptotic
behaviour of the line at criticality was found to be simply that of a
free line ($P({\bf 0},n\tau) \sim 1/n$). In the bound phase the line
was found to be essentially bound to the attractive defects, the
density at the repulsive defects being smaller by a factor of
$(1/l^{*})^{2}$.

We feel that the introduction of sparse potentials introduces some
simplifying features into the study of directed lines. In the current
paper we have examined probably the simplest form for these
potentials; namely single defects, and periodic arrays of defects;
although even for these simple periodic arrays there are many more
features that may be examined, such as the spatial variation of the
probability density away from the column, and the behaviour of the
system in dimensions other than (2+1). One may also retain the
simplifying nature of a periodic array of sparse potentials, but relax
the condition used in this paper that the potentials are short-ranged
in the transverse directions. For instance it would be of interest to
study potentials which were periodic in the transverse dimensions, as
one may then make contact to systems in which a directed line is
interacting with a set of crystal layer potentials. Part of our
motivation for examining sparse potentials was to see if analytic
progress is possible for simple types of disorder - such as randomly
placed defects, or regularly placed defects with random energy. We
certainly regard such analyses as worthwhile and possible projects for
the future.

\vspace{5mm}

We are grateful to Alan Bray and an anonymous referee for useful
comments. TJN acknowledges financial support from the Engineering and
Physical Science Research Council.

\newpage

\appendix

\section{}
In this appendix we outline the solution of the recurrence relation
for the partition function evaluated at the defect sites, $\psi _{n}$.
This is given by (\ref{final1}):
\begin{equation}
\label{app1}
\psi _{n} = f_{n} + R \sum \limits _{m=1}^{n-1} f_{n-m} \psi _{m} \ .
\end{equation}
The solution is most easily obtained by introducing the generating
function
\begin{equation}
\label{gen}
{\bar \psi}(z) \equiv \sum \limits _{n=1}^{\infty} z^{n} \psi _{n} \ ,
\end{equation}
along with a similar function ${\bar f}(z)$ defined in terms of
$\lbrace f_{n} \rbrace $.  Summing (\ref{app1}) over $n$ with the
appropriate weight then gives
\begin{equation}
\label{app2}
{\bar \psi}(z) = {{\bar f}(z) \over [ 1 - R {\bar f}(z) ] } \ .
\end{equation}
Inverting the relation (\ref{gen}) using the calculus of residues then
yields the solution
\begin{equation}
\label{sola}
\psi _{n} = {1 \over 2\pi i} \ \oint \limits _{C_{1}} {dz \over
  z^{n+1}} {\bar \psi }(z) = {1 \over 2\pi i R} \ \oint \limits
_{C_{1}} {dz \over z^{n+1}} \ {1 \over \lbrack 1 - R {\bar f}(z)
  \rbrack } \ ,
\end{equation}
where the contour $C_{1}$ is a closed circle of radius $\delta $; this
circle being chosen so that no other singularities are enclosed bar
the $n ^{\rm th}$ order pole at the origin.

The function ${\bar f}(z)$ is easily evaluated to be
\begin{equation}
\label{fbar}
{\bar f}(z) = (\pi l^{2})^{-1} \ \sum \limits _{n=1}^{\infty} \frac
{z^{n}}{n} = -(\pi l^{2})^{-1} \ \ln (1-z) \ ,
\end{equation}
where the sum is guaranteed to converge since $|z| < \delta \ll 1$.
It is the ease with which this sum may be evaluated in
$(2+1)$-dimensions which makes this case the most analytically
tractable.  Defining $R ' = R / (\pi l^{2})$, we have the explicit
form for $\psi _{n}$ as
\begin{equation}
\label{solb}
\psi _{n} = {1 \over 2\pi i R } \ \oint \limits _{C_{1}} {dz \over
  z^{n+1}} \ {1 \over \lbrack 1 + R ' \ln (1-z) \rbrack } \ .
\end{equation}
Examination of the integrand reveals that there exist two
singularities in the complex plane apart from the pole at the origin.
These are a branch point at $z=1$ along with a simple pole at
\begin{equation}
\label{polepos}
z_{p} = 1 - \exp (-1/R') = 1 - \exp \left\{ -\frac{(l^{*})^2} {e^{\rho
    ^{*}} -1} \right\} \ .
\end{equation}
By cutting the contour $C_{1}$ on the negative real axis, we may wrap
it around the rest of the complex plane as illustrated in fig. A1. In
this way, we have
$$ \oint _{C_1} \ + \ \int _{\rm cut}\ + \ {\rm residue \ at} \ z_{p}
\ = \ 0 \ .$$ We have thus replaced the essentially perturbative
expression (\ref{solb}), by an expression which enables us to extract
the strong-coupling behaviour, if it exists (this actually depends on
the existence of a pole at radius $\delta < |z| < 1$.)

Whether the residue from the pole at $z_{p}$ dominates over the
contribution from the branch cut, depends on the value of $|z_{p}|$.
It is clear from the form of the integrand that the integral will be
dominated (for large $n$) by the pole if it lies within the unit
circle. From (\ref{polepos}) it follows that $z_{p} < 1 - \exp
(l^{*})^2$ if the defects are repulsive and $0<z_{p}<1$ if the defects
are attractive. Thus for attractive defects the pole dominates and a
calculation of the residue leads directly to (\ref{asym1}).

In the case of repulsive defects, the pole is seen to lie on the
negative real axis far outside the circle $|z|=1$. Thus the residue at
$z_{p}$ yields only an exponentially decaying contribution and is
sub-dominant to the contribution from the cut. We therefore have
$\oint _{C_1} \sim -\int _{\rm cut}$, which has the explicit form
\begin{equation}
\label{cutrep}
\psi _{n} \sim {1 \over \pi l^{2} } \int \limits _{0}^{\infty} {dx
  \over (1+x)^{n+1}} \ \left \lbrace {1 \over (\pi R ')^{2} + (1 + R '
  \ln x )^{2} }\right \rbrace \ .
\end{equation} 
Referring to (\ref{norm1}) we may express the normalization in
integral form.  Explicitly one finds
\begin{equation}
\label{cutrepn}
N _{n} \sim 1 + R ' \int \limits _{0}^{\infty} {dx \over x} \ \left
\lbrace (1+x)^{n} - 1 \over (1+x)^{n+1} \right \rbrace \ \left \lbrace
{1 \over (\pi R ')^{2} + (1 + R ' \ln x )^{2} }\right \rbrace \ .
\end{equation} 

We shall briefly describe the asymptotic evaluation of (\ref{cutrep}).
The integral form of the normalization (as well as the similar
integrals which appear in the evaluation of $P({\bf 0},t)$ for $t \ne
n\tau $) may be done in analogous fashion. So referring to
(\ref{cutrep}), as a first step we scale $x$ by $n$, and use the
relation $\exp(p) = \lim _{n \rightarrow \infty} (1+p/n)^{n}$. We then
have
\begin{equation}
\label{cutrep2}
\psi _{n} \sim {1 \over n \pi l^{2}(R ')^{2} } \int \limits
_{0}^{\infty} dx \ e^{-x} \ \left \lbrace {1 \over \pi ^{2} + \lbrack
  \ln (x /\beta) \rbrack ^{2} }\right \rbrace \ ,
\end{equation} 
where $\beta = ne^{- 1/ R '} > ne^{(l^{*})^{2}} \gg 1$. We now split
the integration range into three regions and estimate the order of
magnitude of the integral in each region. Region i) is defined by $0 <
x < 1/\beta $, and retaining only dominant terms for small $x$, we
find (up to prefactors) $\psi _{n}^{(i)} \sim ( \beta \ln
^{2}(\beta))^{-1}$.  Region ii) is defined by $1/\beta < x < \beta$.
In this region we may drop $\ln(x)$ in comparison to $\ln(\beta )$
which gives $\psi _{n}^{(ii)} \sim ( \ln ^{2}(\beta))^{-1}$.  Region
iii) is defined by $x > \beta $ in which case $\psi _{n}^{iii} \sim
e^{-\beta }$.  So clearly the contribution from region ii) dominates
for large $n$. In a similar way, one may establish that the
normalization has the asymptotic form of $N_{n} \sim O(1) +
O(1/\ln(n))$. Putting these results together gives the form of the
probability density shown in (\ref{probrep}).

\section{}
In this appendix we outline the solution of the simultaneous iterative
equations (\ref{iteo}). As before, it is convenient to use generating
functions. Thus we define the functions
\begin{equation}
\label{gen1}
{\bar \psi}^{E}(z) \equiv \sum \limits _{n=1}^{\infty} z^{2n} \psi
^{E}_{n} \ , \ \ \ \ {\bar \psi}^{O}(z) \equiv \sum \limits
_{n=1}^{\infty} z^{2n-1} \psi ^{O}_{n} \ .
\end{equation}
We also define
\begin{equation}
\label{pro1}
{\bar f}^{E}(z) \equiv \sum \limits _{n=1}^{\infty} z^{2n} f_{2n} \ ,
\ \ \ \ {\bar f}^{O}(z) \equiv \sum \limits _{n=1}^{\infty} z^{2n-1}
f_{2n-1} \ .
\end{equation}
Summing the iterative equations over the appropriate weight and using
the above definitions yields the algebraic equations
\begin{eqnarray}
\label{alge}
{\bar \psi}^{E}(z) & = & {\bar f}^{E}(z) + R {\bar f}^{E}(z) {\bar
  \psi}^{E}(z) - S {\bar f}^{O}(z){\bar \psi}^{O}(z) \\ {\bar
  \psi}^{O}(z) & = & {\bar f}^{O}(z) + R {\bar f}^{O}(z) {\bar
  \psi}^{E}(z) - S {\bar f}^{E}(z){\bar \psi}^{O}(z) \ ,
\end{eqnarray}
which may be readily solved, yielding the solutions
\begin{equation}
\label{solu1}
{\bar \psi }^{E}(z) = {{\bar f}^{E}(z) + S [{\bar f}^{E}(z)^{2} -{\bar
    f}^{O}(z)^{2} ] \over 1 - (R - S) {\bar f}^{E}(z) -R S [{\bar
    f}^{E}(z)^{2}-{\bar f}^{O}(z)^{2}] } \ ,
\end{equation}
and
\begin{equation}
\label{solu2}
{\bar \psi }^{O}(z) = {{\bar f}^{O}(z) \over 1 - (R - S) {\bar
    f}^{E}(z) -R S [{\bar f}^{E}(z)^{2}-{\bar f}^{O}(z)^{2}] } \ .
\end{equation}
From the particular form of $f_{n}$, we also have
\begin{equation}
\label{even}
{\bar f}^{E}(z) = -{1 \over 2\pi l^{2}} \ \ln (1-z^{2}) \ ,
\end{equation}
and
\begin{equation}
\label{odd}
{\bar f}^{O}(z) = -{1 \over 2\pi l^{2}} \ \ln \left ( {1-z \over 1+z}
\right ) \ .
\end{equation}

Given the definition of the generating functions, we may retrieve the
original partition functions using
\begin{equation}
\label{sola1}
\psi _{n}^{E} = {1 \over 2\pi i} \ \oint \limits _{C_{1}} {dz \over
  z^{2n+1}} \ {\bar \psi}^{E}(z) \ , \ \ \ \ \psi _{n}^{O} = {1 \over
  2\pi i} \ \oint \limits _{C_{1}} {dz \over z^{2n}} \ {\bar
  \psi}^{O}(z) \ ,
\end{equation}
where as before $C_{1}$ is a circle enclosing the origin of small
enough radius such that it encloses no singularities other than the
pole at the origin.

We now examine the singularity structure in the complex $z$ plane.
Firstly we note that there now exist two branch points, at $z=\pm 1$
which we connect to infinity with cuts along the real axis as shown in
fig. B1. Also, any pole which may exist within the unit circle will
have a twin reflected through the origin due to fact that the
denominators of the generating functions are even functions of $z$. As
before we replace the perturbative expression by the non-perturbative
one by cutting the contour $C_{1}$ and extending around the
singularities in the complex plane. This is illustrated in fig. B1 and
leads us to the expression
$$ \oint _{C_1} \ + \ \int _{\rm - cut } \ + \ \int _{\rm + cut} \ + \ 
\sum \limits _{z=\pm z_{p}} {\rm residues} \ = \ 0 \ .$$

The existence of a bound state will arise only from there being a pole
within the unit circle. We thus examine the zeroes of the denominator
of the generating functions -- i.e.  the positions of the poles are
solutions of
\begin{equation}
\label{poless}
1 + \frac{1}{2}(R ' - S ' )\ln (1-z^{2}) - R ' S ' \ln (1+z) \ln (1-z)
= 0 \ ,
\end{equation}
where we have defined $R ' = R/(\pi l^{2})$, as before, and similarly
$S ' = S/(\pi l^{2})$. We analyse this equation by first noting that
if $\rho$ grows with $a$ more slowly than $a^2$, in particular if
$\rho = O(1)$, then there is no solution for any $\sigma$. On the
other hand, if $\rho$ grows like $a^2$, or faster, then so do $R$ and
$S$ and the final term in (\ref{poless}) plays only a sub-dominant
role. Therefore the poles are situated at $\pm z_{p}$, with
\begin{equation}
\label{positions}
z_{p} = \left \lbrack 1 - \exp \left ( - {2 \over (R ' - S ' ) }
\right ) \right \rbrack ^{1/2} \ .
\end{equation}
As we vary the parameters $\rho $ and $\sigma $, the poles exist
within the unit circle only when $R ' > S '$. Therefore there is a
critical line in the $(\rho , \sigma )$ plane which separates the
region where poles exist from the region where they do not, which has
the equation $R' = S'$. A more precise equation for this line may be
obtained by including the final term in (\ref{poless}) in the analysis
by substituting $z_{p} = 1 - \epsilon$ into the equation and solving
it in the limit $\epsilon \rightarrow 0$. The resulting condition for
criticality is now
\begin{equation}
\label{crit}
S ' = {R ' \over 1 + 2\ln 2 \, R ' }\ .
\end{equation}
The right-hand-side is a monotonically increasing function of $R'$,
which reaches the value $(l^{*})^{-2}$ (corresponding to $\sigma
\rightarrow \infty$) when $\rho = \pi a^{2}\ln 2 \equiv \rho _{c}$.
The general features of the critical line are now easy to find: a
linear regime near the origin and a logarithmic approach to $\rho
_{c}$ from below. This behavior is summarized in (\ref{critcond}) and
illustrated in fig.1.

In order to examine the behaviour of the line at criticality, we
insert the critical condition (\ref{crit}) into the expressions for
the generating functions. One then finds that to leading order ${\bar
  \psi}^{E}(z) = {\bar f}^{E}(z)$ and ${\bar \psi}^{O}(z) = {\bar
  f}^{O}(z)$, which directly gives the partition function at the
defect sites with no integration required; i.e. we have
\begin{equation}
\label{partition}
\psi _{n}^{E} = 1/(2n \pi l^{2}) \ , \ \ \ \psi _{n}^{O} = 1/((2n-1)
\pi l^{2}) \ .
\end{equation}

The appropriate normalization factors may be derived in an analogous
way to that described in section 5. Explicitly one defines
\begin{equation}
\label{norms1}
N^{E}_{n} = \lim _{\epsilon \rightarrow 0} \int d^{d}x \ Z({\bf x},
2n\tau - \epsilon) \ ,
\end{equation}
and
\begin{equation}
\label{norms2}
N^{O}_{n} = \lim _{\epsilon \rightarrow 0} \int d^{d}x \ Z({\bf x},
(2n-1)\tau - \epsilon) \ .
\end{equation}
These functions then satisfy the relations $N^{E}_{n} = N^{O}_{n}$,
and
\begin{equation}
\label{norr}
N^{E}_{n} = 1 + R \sum \limits _{m=1}^{n-1} \psi _{m}^{E} - S \sum
\limits _{m=1}^{n} \psi _{m}^{O} \ .
\end{equation}
At criticality these functions are asymptotically constants which
along with the results for the partition functions (\ref{partition})
leads to the asymptotic form of the probability density following
$P({\bf 0}, n\tau ) \sim 1/n$.

In order to examine the bound state one may simply ignore the
sub-dominant contributions from the cuts, and evaluate the residues of
the poles at $\pm z_{p}$. No explicit details of this calculation are
given here as it may easily be reconstructed from the analogous case
examined in the previous appendix.

\newpage

{\bf Figure Captions}

\vspace{20mm}

\noindent
Fig.~1: The phase diagram in the $(\sigma, \rho )$ plane. The
critical line separating unbound and bound phases in given
by (\ref{critcond}).

\vspace{10mm}

\noindent
Fig.~A1: The singularity structure in the complex $z$ plane
for the evaluation of ${\bar \psi}(z)$. Also illustrated is
the deformation (solid line) of the original contour $C_{1}$
(dashed line.)

\vspace{10mm}

\noindent
Fig.~B1: The singularity structure in the complex $z$ plane
for the evaluation of ${\bar \psi}^{E}(z)$ and ${\bar \psi}^{O}(z)$. 
Also illustrated is the deformation (solid line) of the original 
contour $C_{1}$ (dashed line.)

\newpage

\vglue 2in

\begin{center}
\leavevmode
\hbox{%
\epsfxsize=450pt
\epsffile{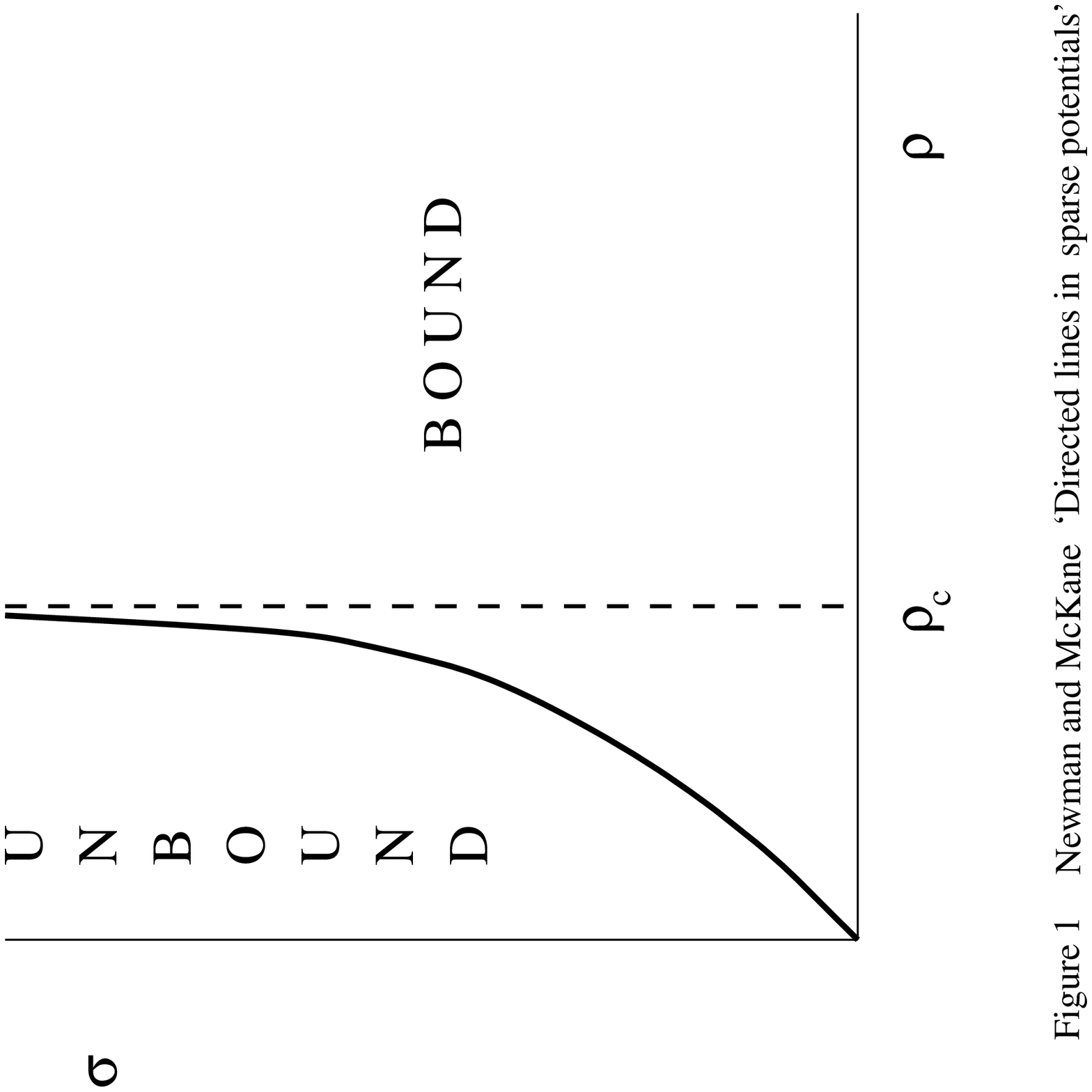}}
\end{center}

\newpage

\begin{center}
\leavevmode
\hbox{%
\epsfxsize=400pt
\epsffile{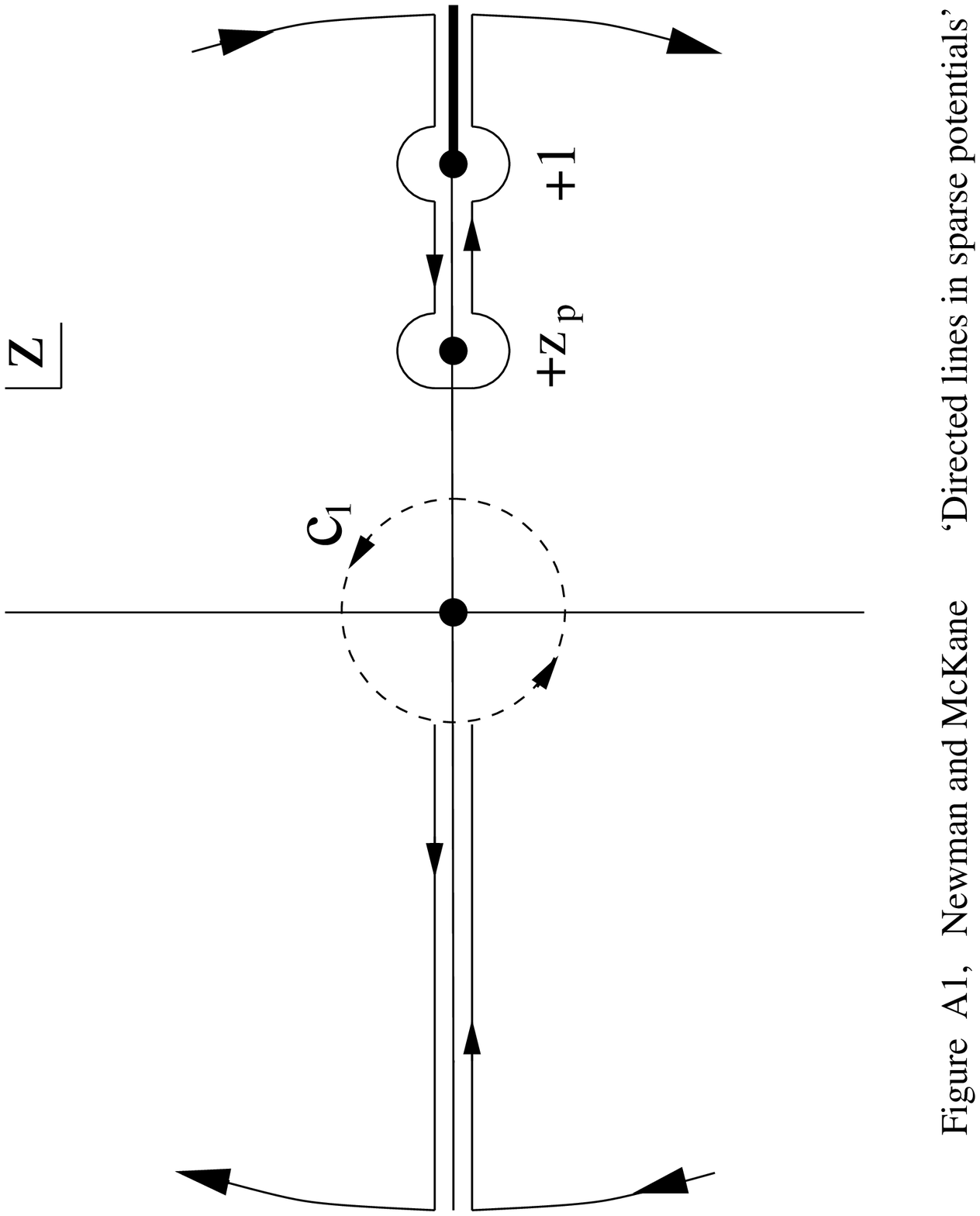}}
\end{center}

\vspace{20mm}

\begin{center}
\leavevmode
\hbox{%
\epsfxsize=400pt
\epsffile{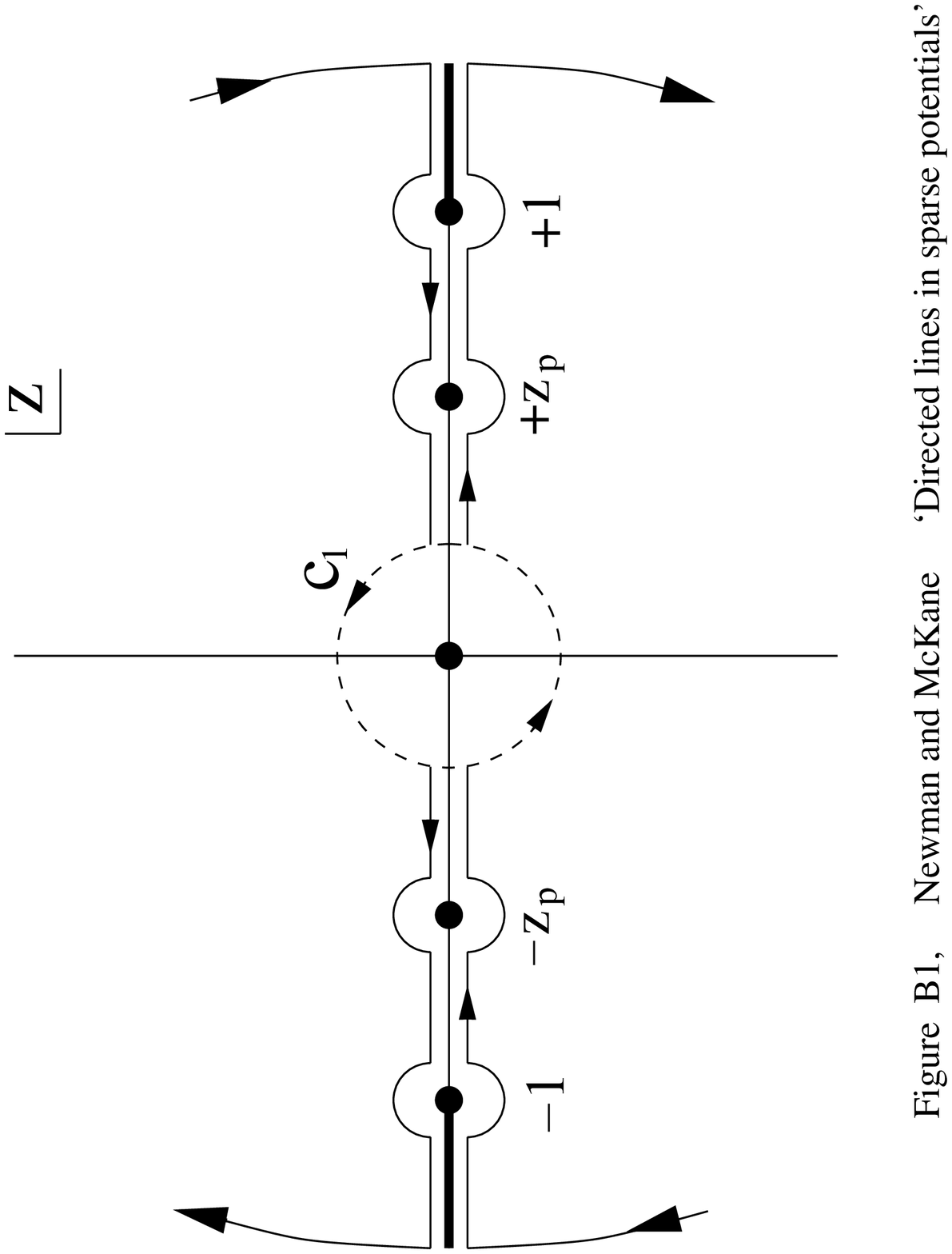}}
\end{center}

\end{document}